# Effect of electron-electron scattering on magnetointersubband resistance oscillations of two-dimensional electrons in GaAs quantum wells

A. V. Goran, A. A. Bykov, and A. I. Toropov
*Institute of Semiconductor Physics, 630090 Novosibirsk, Russia*

S. A. Vitkalov
*Physics Department, City College of the City University of New York, New York 10031, USA*

The low-temperature(4.2 < $T$ < 12.5 K) magnetotransport ($B$ < 2 T) of two-dimensional electrons occupying two subbands (with energy $E_1$ and $E_2$) is investigated in GaAs single quantum well with AlAs/GaAs superlattice barriers. Two series of Shubnikov-de Haas oscillations are found to be accompanied by magnetointersubband (MIS) oscillations, periodic in the inverse magnetic field. The period of the MIS oscillations obeys condition $\Delta_{12}=(E_2-E_1)=k \cdot \hbar w_c$, where $\Delta_{12}$ is the subband energy separation, $w_c$ is the cyclotron frequency, and $k$ is the positive integer. At $T$=4.2 K the oscillations manifest themselves up to $k$=100. Strong temperature suppression of the magnetointersubband oscillations is observed. We show that the suppression is a result of electron-electron scattering. Our results are in good agreement with recent experiments, indicating that the sensitivity to electron-electron interaction is the fundamental property of magnetoresistance oscillations, originating from the second-order Dingle factor.

The Landau quantization in quasi-two-dimensional (2D) electron systems (with two or more occupied subbands) manifests itself in two or more sets of Landau levels. Resonance transitions of electrons between Landau levels corresponding to different two-dimensional subbands [1,2] causes the so-called magnetointersubband (MIS) oscillations of the resistance $r_{xx}$ [3–5]. The interaction between two subbands can be also significant for other phenomena such as cyclotron resonance [6]. The position of the maxima of the MIS oscillations obeys the condition $\Delta_{12}=E_2-E_1=k \cdot \hbar w_c$, where $\Delta_{12}$ is the intersubband energy gap, $E_i$ is the energy of the bottom of $i$th subband, $w_c$ is the cyclotron frequency, and index $k$ is the positive integer. The oscillations, similar to well-known Shubnikov-de Haas (SdH) oscillations, are periodic in the inverse magnetic field and appear in classically strong magnetic fields. The amplitude of SdH oscillations is limited by the broadening of Landau levels due to scattering and by thermal broadening of the Fermi distribution. With increasing temperature the thermal broadening of the Fermi distribution becomes the dominant factor, limiting the amplitude of SdH oscillations. MIS oscillations are significantly less sensitive to the electron distribution and their amplitude is predominantly determined by a quantum relaxation time $t_q$ [4,5].

MIS oscillations were recently observed in GaAs double quantum wells with AlAs/GaAs superlattice barriers with roughly equal electron densities in subbands ($n_1 \approx n_2$) [7–10]. The quantum lifetimes of the electrons in subbands was also approximately equal ($t_{q1} \approx t_{q2}$) [11]. In the general case of two populated subbands the amplitude of the MIS oscillations of the resistance $\Delta r_{MISO}$ depends on the sum of the quantum scattering rates in each subband [4,5]

$$\Delta r_{MISO} = \frac{2m}{e^2(n_1+n_2)} \cdot n_{12} \cdot \exp[-(p/w_c)(1/t_{q1}+1/t_{q2})] \cdot \cos(\frac{2p\Delta_{12}}{\hbar w_c}), \quad (1)$$

where $1/t_{qi}$ and $n_i$ are the quantum scattering rate and electron density in $i$th subband, and $m$ is electron band mass. Parameter $n_{12}$ is an effective intersubband scattering rate [5].

In this Brief Report we investigate MIS oscillations in GaAs single quantum well with AlAs/GaAs superlattice barriers, when two subbands are occupied with substantially different densities. In the temperature range of $T$=4.2–12.5 K a strong decrease in the amplitude of the MIS oscillations with the temperature is observed. We consider the suppression of the MIS oscillations as a result of a temperature dependence on quantum scattering times $t_{qi}$. We have found that the temperature variation in the sum of quantum scattering rates ($1/t_{q1}+1/t_{q2}$) is proportional to $T^2$, indicating dominant contribution of the electron-electron scattering to the suppression of the MIS oscillations.

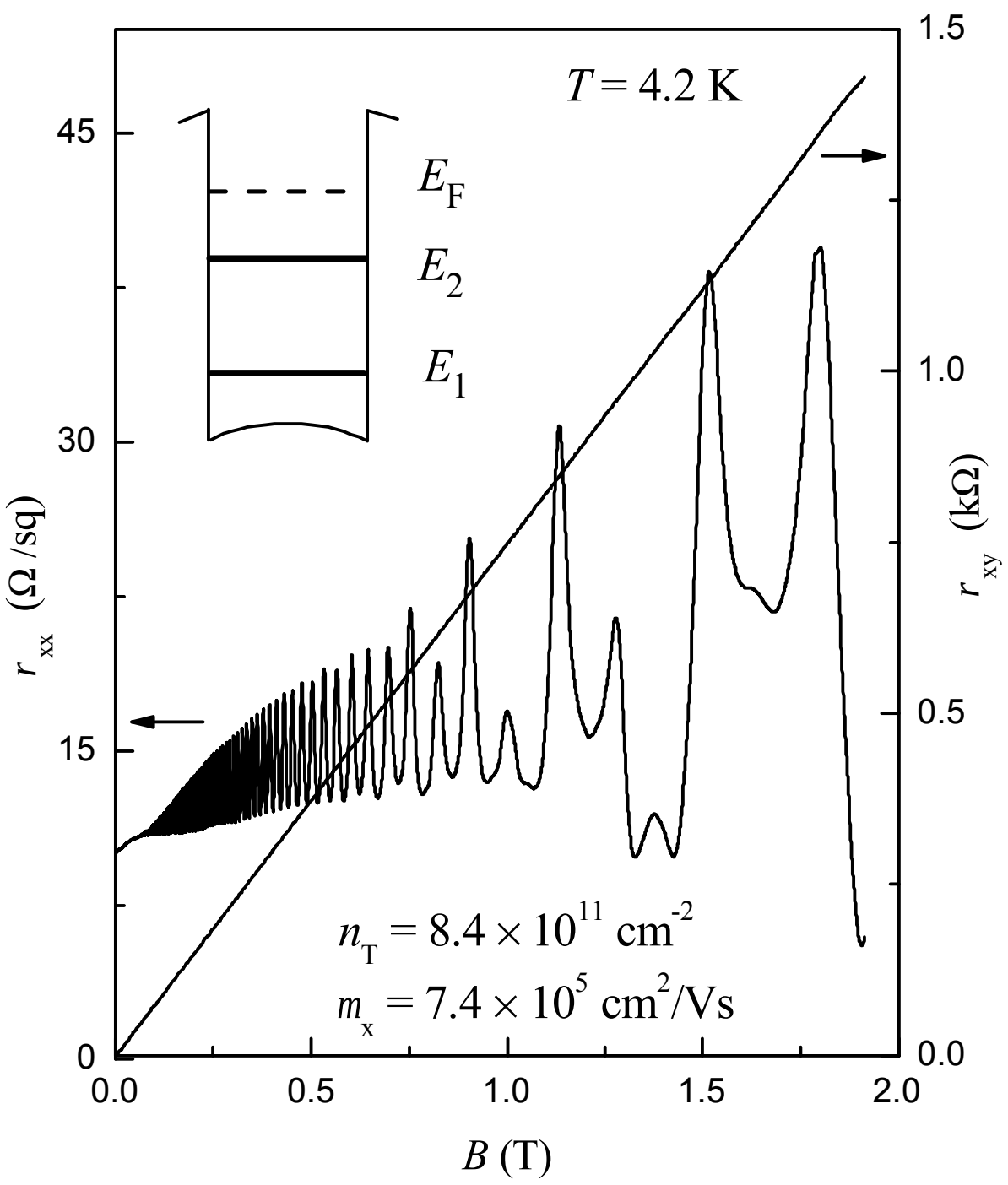

FIG. 1. Resistance $r_{xy}(B)$ and $r_{xx}(B)$ in GaAs quantum well with AlAs/GaAs superlattice barriers at $T$=4.2 K. The inset shows the diagram of the quantum well with two occupied subbands with energies at the bottom of the subbands $E_1$ and $E_2$.

Heterostructures under the study were symmetrically doped GaAs single quantum wells with a width of 26 nm and AlAs/GaAs superlattice barriers [12,13]. The diagram of the quantum well with two occupied subbands $E_1$ and $E_2$ is presented in the inset of Fig. 1. The structures were grown on GaAs substrates, whose deviation from the (100) plane did not exceed 0.02°. The measurements were carried out in the temperature range of $T$=4.2–12.5 K in the magnetic field $B$<2 T on 450×50 $\mu$m Hall bars, fabricated using optical lithography and liquid etching. Magnetoresistance $r_{xx}(B)$ and $r_{xy}(B)$ was measured using ac current $I_{ac}$<1 μA in the frequency range of 0.01–1 kHz. The total electron density, $n_T$=8.4·10$^{11}$ cm$^{-2}$, was calculated from the Hall resistance $r_{xy}$ in magnetic field $B$=0.5 T. The electron mobility $m_x$=7.4·10$^5$ cm$^2$/ Vs was obtained from $n_T$ and zero-field resistance $r_{xx}(B=0)=r_0$ at liquid-helium temperature.

Figure 1 presents typical experimental curves of resistance $r_{xy}(B)$ and $r_{xx}(B)$ in studied samples. In magnetic field $B$<1 T the Hall resistance $r_{xy}(B)$ follows a straight line, indicating negligible contributions of the Landau quantization to the response. The slope of $r_{xy}(B)$ depends on the total electron density $n_T$. At $T$=4.2 K the longitudinal resistance $r_{xx}$ oscillates with magnetic fields at $B$>0.07 T. For the quantum well with two occupied subbands the oscillatory part of $r_{xx}(B)$ contains two series of SdH oscillations and the magnetointersubband oscillations [1–5], dominating at low magnetic fields $B$<0.5 T.

Figure 2 presents $r_{xx}(1/B)$ dependence, demonstrating, that $r_{xx}$ oscillations have only one period at the low magnetic fields ($1/B > 2$ T$^{-1}$). The Fourier transform of the oscillatory part of $r_{xx}(1/B)$ is shown in the inset of Fig. 2. Three frequencies, associated with the peaks and marked as $f_1$, $f_2$, and $f_3$, are identified as follows. The frequency $f_3$ corresponds to magnetointersubband oscillations, while $f_1$ =13 T$^{-1}$ and $f_2$=4 T$^{-1}$ correspond to two series of SdH oscillations from the two subbands with electron densities $n_1=2ef_1/h$=6.3·10$^{11}$ cm$^{-2}$ and $n_2=2ef_2/h$=1.9·10$^{11}$ cm$^{-2}$. The frequency of the highest peak $f_3$=9.1 T$^{-1}$ equals (with an accuracy of 1%) to the difference between the other two frequencies $f_1-f_2$.

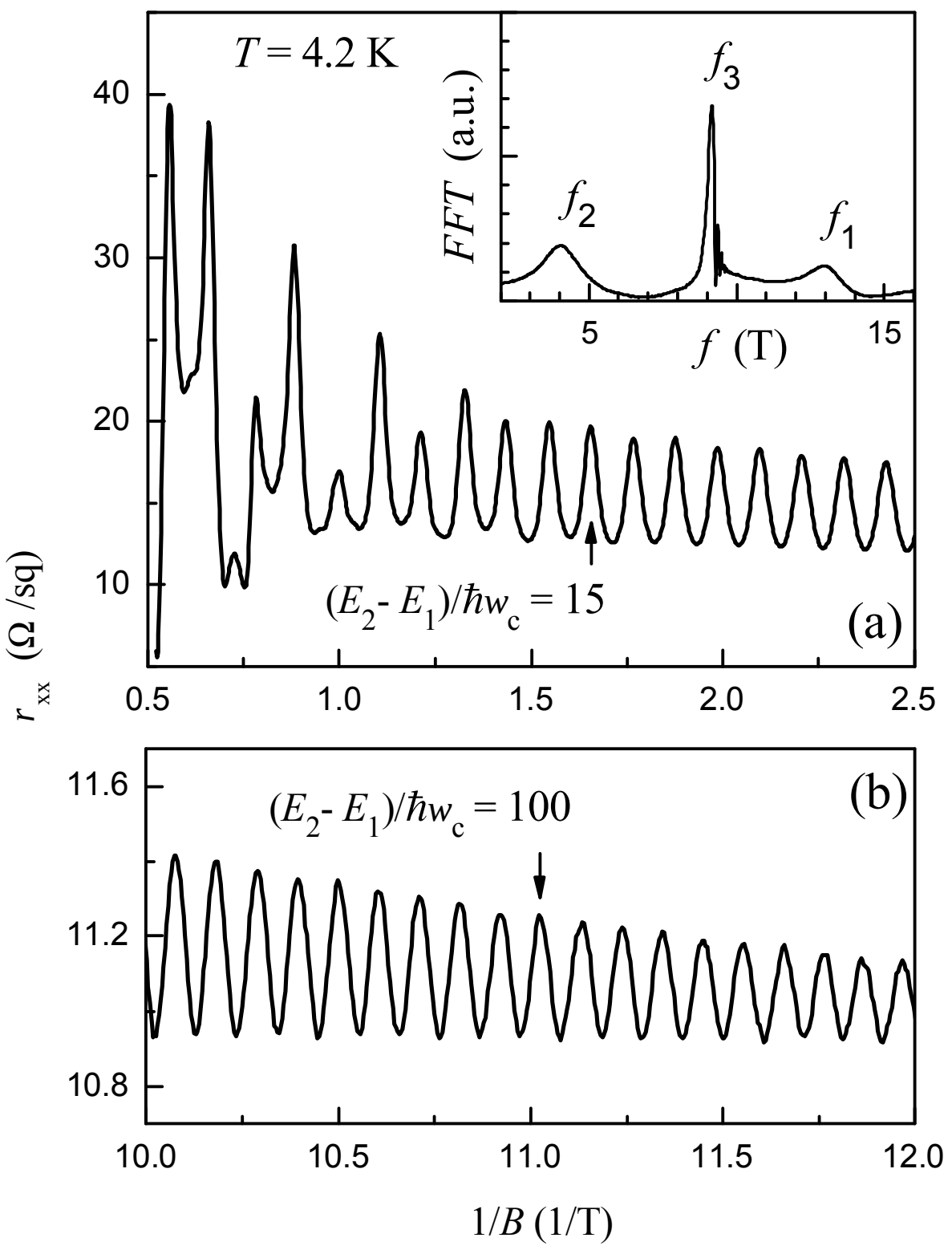

FIG. 2. Resistance $r_{xx}$ vs inverse magnetic field $1/B$ in GaAs quantum well with two occupied subbands $E_1$ and $E_2$ at $T$=4.2 K. Arrows mark the positions of the maxima corresponding to $k=\Delta_{12}/\hbar w_c$ =15 and 100. The inset shows the Fourier transform of the oscillatory part of the resistance $r_{xx}$ $(1/B)$.

The sum of the two densities ($n_1+n_2=8.2\cdot10^{11}$ cm$^{-2}$) is, to a high degree of accuracy (2%), equal to the total density $n_T$, calculated from the Hall resistance. The intersubband energy gap $\Delta_{12}$=15.8 meV, calculated from the frequency $f_3$, is in a good agreement with band-structure calculations of the studied GaAs quantum well. Thus, the experimental data indicate that the magnetoresistance oscillations $r_{xx}$ with frequency $f_3$ are the magnetointersubband oscillations and the position of the oscillation maxima obeys the relation $\Delta_{12}=k\cdot\hbar w_c$, where $\Delta_{12}$=15.8 meV and $k$ is the positive integer. It is worth mentioning, that at helium temperatures our samples show the MIS oscillations with the index $k$ up to 100, indicating a very high quality of our samples. It has provided the study of the oscillations in a very broad range of the inverse magnetic fields.

Figure 3(a) presents the experimental dependences of $r_{xx}/r_0(1/B)$ at temperatures of 4.2 and 12.5 K. One can see a substantial suppression of MIS oscillations with the increase of the temperature. Figure 3(b) shows the amplitudes of the magnetointersubband oscillations plotted against $1/B$ on a semilogarithmic scale. In a wide range of magnetic fields the amplitude of the oscillations are well approximated by the expression [Eq. (1)] with a constant ($B$-independent) intersubband scattering rate $n_{12}$ and the $B$-independent broadening of Landau levels ($\sim1/t_{qi}$). At $T$=4.2 K and in very small magnetic fields $B$<0.1 T the approximation fails. Reasons of the discrepancy are not clear at the moment. A failure of the linear approximation is also observable at $T$=12.5 K at high magnetic fields. We attribute this discrepancy to the influence of magnetophonon resonance seen in high-mobility semiconductor structures at large filling factors [14,15].

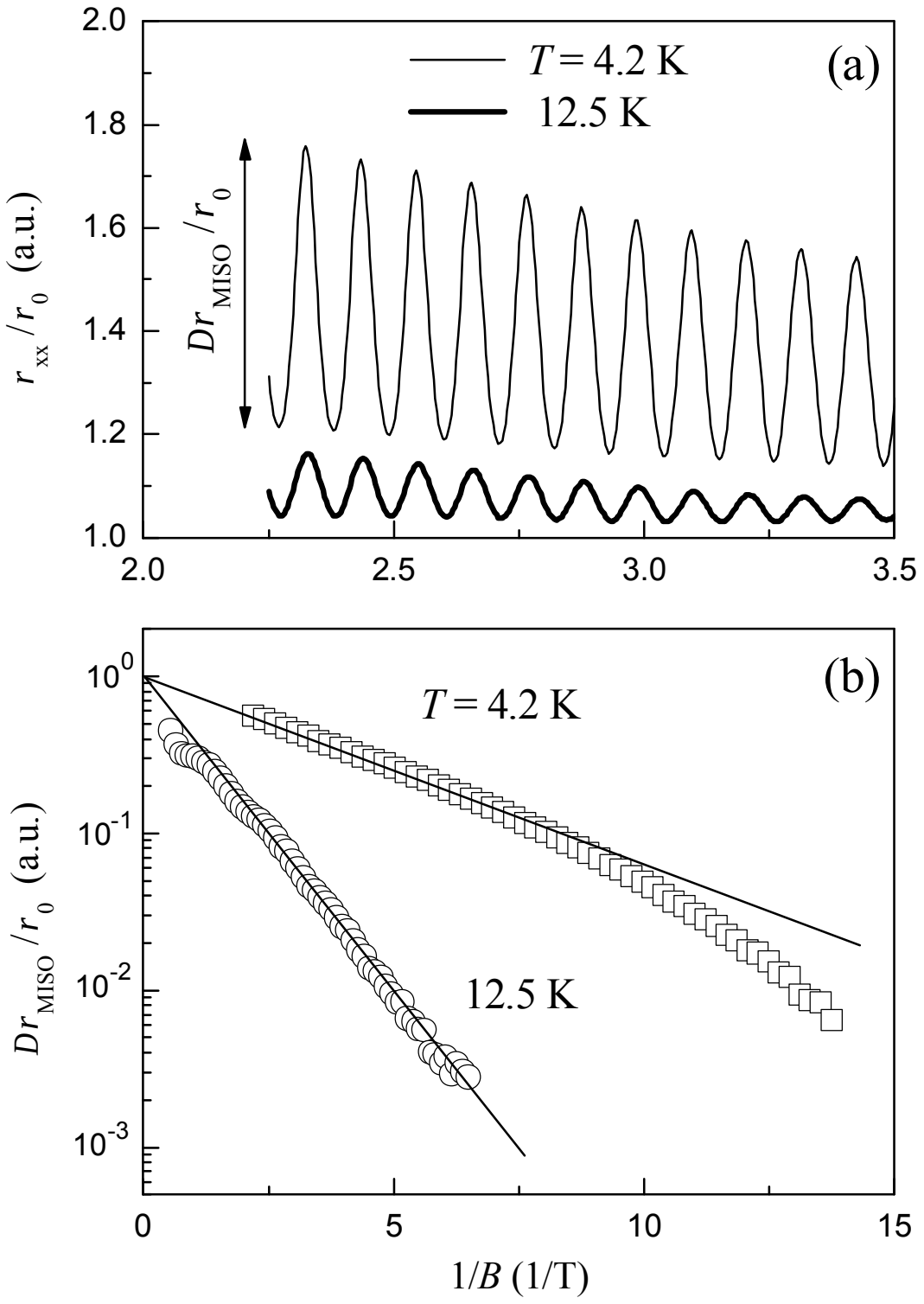

FIG. 3. (a) Normalized $r_{xx}/r_0$ vs $1/B$ in GaAs quantum well with two occupied subbands at temperatures 4.2 and 12.5 K. (b) $\Delta r_{MISO}/r_0$ vs $1/B$ at $T$=4.2 and 12.5 K. Straight lines correspond to Eq. (1) $\Delta r_{MISO}/r_0=\exp[-(\pi/\omega_c)\cdot(1/t_{q1}+1/t_{q2})]$.

Shown in Fig. 3(a), the strong temperature dependence of MIS oscillations indicates a significant variation in the quantum lifetime of electrons $t_q$ with the temperature. Figure 4 presents the temperature dependence of the sum of quantum scattering rates $1/t_{q1}+1/t_{q2}$. In the plot each point is extracted from the slope of the dependence of the logarithm of the amplitude of MIS oscillations on the inverse magnetic field, measured at different temperatures [see Fig. 3(b)]. In accordance with Eq. (1), the slope is proportional to the sum $1/t_{q1}+1/t_{q2}$. We have found that the temperature variation in the sum of the quantum scattering rates is well approximated by the square of the temperature. Figure 4 presents the data plotted versus $T^2$. A linear plot $1/t_{q1}+1/t_{q2}=A+B\cdot T^2$ provides a reasonable fit with $A$=0.142 (1/ps) and $B$ =0.00415 [1/(ps K$^2$)].

Figure 4 shows that the temperature-dependent terms of the inelastic-scattering rates are proportional to $T^2$, indicating the dominant contribution of the electron-electron interaction to the broadening of the Landau levels. In accordance with the theory [16,17], the contribution of the electron-electron interaction into the electron lifetime is proportional to $T^2$ and inversely proportional to the electron Fermi energy $E_F$. For a two-band system, we have used the following approximation for the $e$-$e$ scattering rate: $1/t_{ee}=l_{eff}(1/E_{F1}+1/E_{F2})\cdot T^2$, where $E_{Fi}=E_F-E_i$ is the Fermi energy counted from the bottom of the $i$th band. A comparison between the expression and the experiment yields $l_{eff}\approx 2$, which is close to theoretical estimations.

Elastic impurity scattering controls the temperature independent term $A$ of the Dingle factor. The obtained value of the sum of the impurity scattering rates $A=1/t_{q_1}^{im}+1/t_{q_2}^{im}$=0.142 (1/ps) indicates the impurity scattering time $t_q^{im}$>14 (ps) in one or both subbands, demonstrating again the high quality of the measured structures. The inset in Fig. 4 shows temperature dependence of the transport scattering rate $1/t_{tr}$, obtained from the resistance at zero magnetic field. The transport scattering rate depends weakly on the temperature. We note also, that in the studied range of temperatures, the value of the quantum scattering time $t_q$ is found to be considerably less than the transport scattering time $t_{tr}$. The high value of the ratio $t_{tr}/t_q$ >>1 is typical for GaAs quantum wells with AlAs/GaAs superlattice barriers [12].

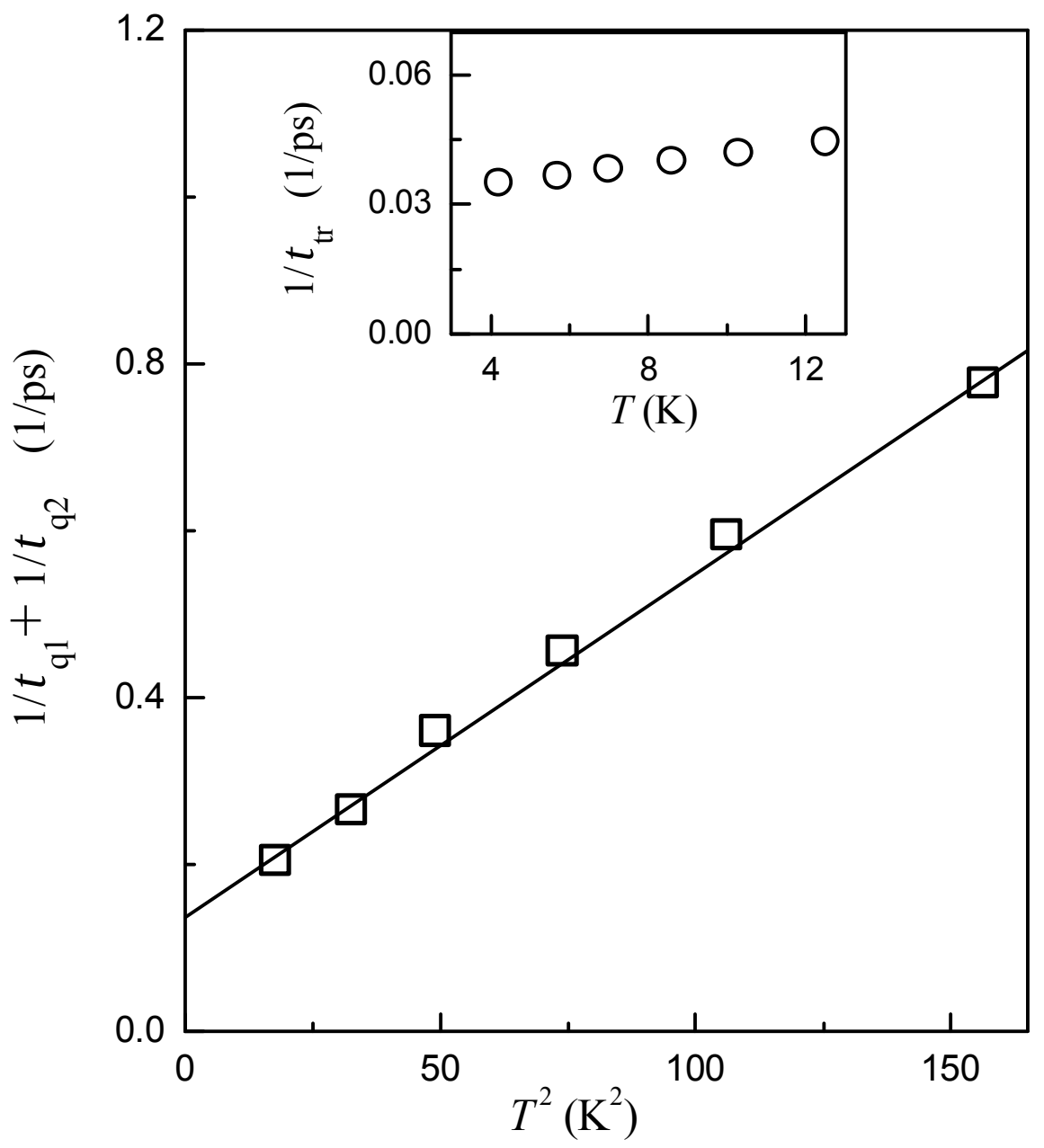

FIG. 4. Sum of the scattering rates $\Sigma_q = 1/t_{q1} + 1/t_{q2}$ vs $T^2$ in GaAs quantum well with two occupied subbands. Straight line corresponds to $\Sigma_q = A + B \cdot T^2$ with $A$=0.142 and $B$=0.00415.

In summary, we have experimentally studied the magnetoresistance of GaAs quantum well with AlAs/GaAs superlattice barriers with two subbands occupied. Magnetointersubband oscillations of the resistance are found with the period, determined by relation $\Delta_{12} = k \cdot \hbar w_c$, where the index $k$ is up to 100 at $T$=4.2 K. The high value of the index $k$ provides measurements of the period with high accuracy, which can be useful for accurate calibration of the magnetic fields. The magnitude of the oscillations depends considerably on temperature, pointing to a large temperature variation in the quantum scattering time of electrons $t_q$. The change in the quantum time is found to be proportional to the square of the temperature. It reveals the electron-electron interaction as the main agent, affecting the electron lifetime $t_q$. Our results are in good agreement with the recent experimental observations [11,18–20], indicating that the sensitivity to electron-electron scattering is the fundamental property of magnetoresistance oscillations originating from the secondorder Dingle factor.

The authors thank L. I. Magarill for useful discussion. The work was supported by RFBR under Project No. 08-02- 01051.